\documentclass[twocolumn,showpacs,preprintnumbers,amsmath,amssymb]{revtex4}

\usepackage{graphicx}
\usepackage{dcolumn}
\usepackage{bm}
\usepackage{color}

\begin{document}


\title{Microwave zero-resistance states in a bilayer electron system}

\author{S. Wiedmann,$^{1,2}$ G. M. Gusev,$^3$ O. E. Raichev,$^4$ A. K. Bakarov,$^5$
and J. C. Portal$^{1,2,6}$} 
 \affiliation{$^1$LNCMI-CNRS, UPR 3228, BP
166, 38042 Grenoble Cedex 9, France} \affiliation{$^2$INSA Toulouse,
31077 Toulouse Cedex 4, France} \affiliation{$^3$Instituto de
F\'{\i}sica da Universidade de S\~ao Paulo, CP 66318 CEP 05315-970,
S\~ao Paulo, SP, Brazil}\affiliation{$^4$Institute of
Semiconductor Physics, NAS of Ukraine, Prospekt Nauki 41, 03028,
Kiev, Ukraine} \affiliation{$^5$Institute of Semiconductor Physics,
Novosibirsk 630090, Russia}\affiliation{$^6$Institut Universitaire
de France, 75005 Paris, France}
\date{\today}

\begin{abstract}
Magnetotransport measurements on a high-mobility electron bilayer system formed in 
a wide GaAs quantum well reveal vanishing dissipative resistance under continuous 
microwave irradiation. Profound zero-resistance states (ZRS) appear even in the 
presence of additional intersubband scattering of electrons. We study the dependence 
of photoresistance on frequency, microwave power. and temperature. Experimental 
results are compared with a theory demonstrating that the conditions for absolute 
negative resistivity correlate with the appearance of ZRS.
\end{abstract}


\maketitle

Exposing two-dimensional electron systems (2DES) to a continuous microwave (MW) irradiation 
in the presence of perpendicular magnetic fields $B$ has revealed notable ``zero-resistance states"
(ZRS) for samples with ultrahigh electron mobility and under high MW intensity \cite{1,2,3}.
The longitudinal resistance $R_{xx}$ vanishes in certain intervals of magnetic fields,
whereas the Hall resistance remains unchanged, i.e., does not show plateaus. The ZRS phenomenon 
is closely related to the microwave-induced resistance oscillations (MIROs) \cite{4} which are 
governed by the ratio of the radiation frequency $\omega$ to the cyclotron frequency $\omega_{c}=
eB/m$, where $m$ is the effective mass of the electrons. If the electron mobility is high enough, 
an increase in MW power transforms the minima of these oscillations to intervals of zero resistance. 

In theory, the effect of microwaves on 2DES is described by two microscopic mechanisms. 
The ``displacement" mechanism is based on spatial displacement 
of electrons along the applied dc field under scattering-assisted microwave absorption \cite{5,6,7}, 
while the ``inelastic" mechanism originates from an oscillatory contribution to the isotropic part of 
the electron distribution function \cite{8,9}, controlled by inelastic relaxation. Both mechanisms 
describe phase and periodicity of MIROs observed in experiments, but for low temperatures $T$ the 
inelastic mechanism is the dominant one because the inelastic relaxation time $\tau_{in}$ is large, 
$\tau_{in} \propto T^{-2}$. According to theory \cite{5,6,7,8}, both mechanisms can lead to negative  
dissipative resistivity in the MIRO minima with increasing MW power. A direct consequence of negative 
resistivity (regardless of its microscopic mechanism) is the instability of homogeneous current 
flow and spontaneous breaking of the sample into a pattern of domains \cite{10}. The simplest 
domain structure of this kind is viewed as two domains with opposite directions of currents and 
Hall fields \cite{7,10}. Each domain is characterized by zero dissipative resistance and classical 
Hall resistance. A change of the current in this regime is accommodated by a shift of the domain 
wall without any voltage drop, so the resistivity becomes zero. More complicated domain structures 
\cite{11} and an alternative mechanism of ZRS, based on the MW stabilization of the edge-state 
transport \cite{12}, have been proposed recently. Until now, all studies of ZRS are restricted to
high-mobility single-layer 2D systems.

Bilayer 2DES, where electrons occupy two closely spaced subbands, exhibit magneto-intersubband 
(MIS) oscillations of magnetoresistance caused by the periodic modulation of the probability of 
intersubband transitions by the magnetic field (see Ref. \cite{13} and references therein). 
In the presence of MW irradiation, the interference of MIROs with MIS oscillations leads to a 
peculiar magnetoresistance picture with enhancement, suppression, or inversion (flip) of MIS peaks 
correlated with MW frequency. The photoresistance in bilayers \cite{14} can be described by the 
inelastic microscopic mechanism at low temperatures. Similar phenomena arise in 
multisubband systems \cite{15}. However, there is neither experimental nor theoretical
evidence for MW-induced ZRS in systems with more than one populated subband. This also raises the 
question of whether ZRS might be observable in quasi-3D systems. 

In this Letter, we present studies of magnetoresistance in a wide quantum well 
(WQW) exposed to MW irradiation. Owing to charge redistribution, WQWs with high electron density 
form a bilayer configuration, where the two wells near the interfaces are separated by an electrostatic 
potential barrier \cite{16}, and two subbands appear as a result of tunnel hybridization 
of 2D electron states. Without microwaves, the magnetoresistance of our system shows 
MIS oscillations, thereby confirming the existence of two occupied subbands. By applying 
microwaves, we have found that for different MW frequencies two of the MIS peaks, which 
are inverted with increasing MW power, evolve into ZRS. Calculations of photoresistance 
demonstrate that ZRS in our samples appear under conditions for absolute negative 
resistivity, which suggests the origin of ZRS as a consequence of current instability.

\begin{figure}[ht]
\includegraphics[width=9cm]{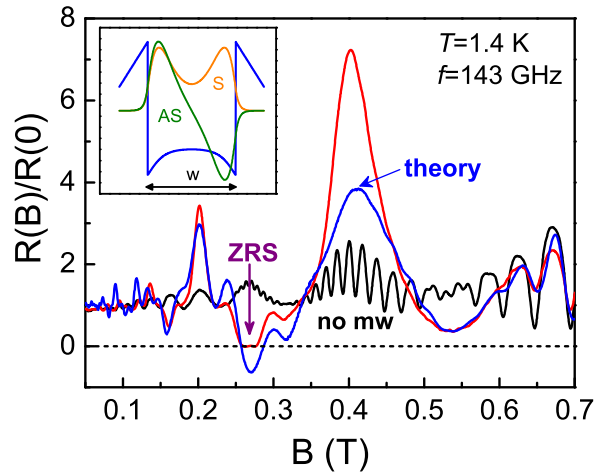}
\caption{\label{fig1} (color online) Longitudinal resistance without (no MW)
and under MW irradiation (143~GHz) at 1.4~K. The microwaves lead to enhancement, 
suppression, or inversion of the MIS peaks. The inverted MIS peak at 0.27~T exhibits 
vanishing resistance. These basic features are also seen in the theoretical plot. 
Inset: WQW with $w=45$~nm and corresponding symmetric (S) 
and asymmetric (AS) wave functions.} 
\end{figure}

We have studied 45~nm wide WQWs with high electron density $n_{s} \simeq 9.1 \times 10^{11}$ cm$^{-2}$, 
obtained from the periodicity of Shubnikov-de Haas (SdH) oscillations, and mobility $\mu~\simeq 
1.9 \times 10^{6}$ cm$^{2}$/V s, at $T=1.4$~K and after a brief illumination with a red light-emitting diode. 
The electrons occupy the two lowest subbands, and the corresponding wave functions, symmetric (S) 
and antisymmetric (AS), are presented in the inset of Fig. \ref{fig1}. The subband separation 
$\Delta$~=~1.40~meV is extracted from the periodicity of low-field MIS oscillations \cite{13} in 
agreement with theoretical calculations. The measurements were carried out in a cryostat with a variable 
temperature insert and a waveguide was employed to deliver linearly polarized MW radiation 
(frequency range 32.7–170 GHz) down to the sample. A conventional lock-in technique was used 
to measure the longitudinal resistance $R=R_{xx}$. The samples have van der Pauw-geometry 
(3~mm~$\times$~3~mm), and the longitudinal resistance is found to be insensitive to the orientation 
of the microwave field.

In Fig. \ref{fig1} we present dark resistance and the observation of a ZRS (marked with an arrow) 
for 143~GHz at a temperature of 1.4~K. Other MIS peaks, starting from $B=0.07$~T, are either 
inverted or enhanced (since $\hbar\omega$ is close to $\Delta/2$, approximately each second peak 
in the region above 0.13~T is inverted) but only the MIS peak at 0.27~T exhibits vanishing 
longitudinal resistance. The resistance without MW irradiation shows MIS oscillations which 
are superimposed on SdH oscillations for $B > 0.25$~T. The SdH oscillations are damped due to 
electron heating by microwaves \cite{14,15}. The minimum at 0.55~T and weak minima at 0.22 and 
0.32 T are not attributed to MIS oscillations. They occur in the regions of transitions from 
inversed to enhanced MIS, also confirmed by theoretical calculations. The theoretical plot 
(see details below) shows a negative resistance in the region of experimentally observed ZRS. 

\begin{figure}[ht]
\includegraphics[width=9cm]{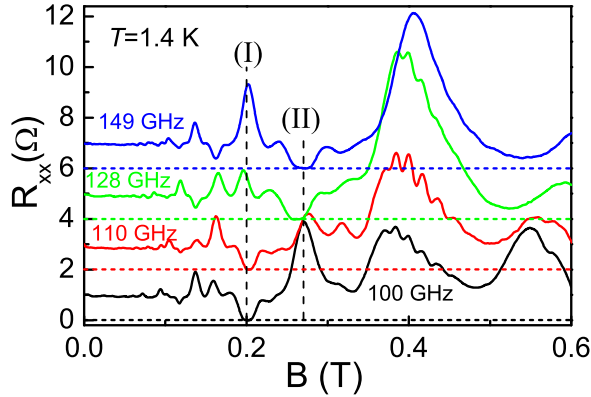}
\caption{\label{fig2} (color online) Frequency-dependent photoresistance for 149, 128, 110, and 
100 GHz exhibits ZRS for two inverted MIS peaks: (I) at 0.2 T and (II) at 0.27 T. Peak (II) is strongly
enhanced for 100 GHz but becomes inverted and shows ZRS for 128 and 149 GHz, while peak (I) is enhanced 
for 149 GHz but shows ZRS for 110 and 100 GHz. Photoresistance traces are shifted up for clarity 
except for 100 GHz.}
\end{figure}

Figure \ref{fig2} illustrates photoresistance in the frequency range from 149~GHz to 
100~GHz with four chosen frequencies at 149, 128, 110, and 100~GHz (from top to bottom). We denote
the ZRS at $B=0.2$~T as peak (I) and the ZRS at $B=0.27$~T as peak (II). For all four traces, we 
show the highest MW intensity (0~dB). As in Fig. \ref{fig1} for 143~GHz, we observe a ZRS for 
peak (II) at 149 and 128~GHz. In contrast to $f \geq 128$~GHz, we find a ZRS for peak (I) 
at 110 and 100~GHz. Note that the deep minima of ZRS slightly shift to lower magnetic fields 
as MW frequency decreases (compare 149 with 128~GHz and 110 with 100~GHz). Peak (I), which shows vanishing 
resistance for 100 and 110~GHz, is strongly enhanced with increasing frequency. The same effect 
occurs for peak (II), which exhibits ZRS for $f>128$~GHz but is first enhanced for 110~GHz 
and strongly enhanced for 100~GHz. It is worth noting that other MIS oscillations at $B< 0.2$~T 
also show enhancement or peak flip correlated with MW frequency but no ZRS. The amplitude of 
MIS peak at $B\simeq 0.4$~T becomes smaller with decreasing frequency. The results in Fig. \ref{fig2} 
demonstrate the striking difference of ZRS in our bilayer system compared to single-subband systems.
In the single-subband systems, MIRO minima evolving into ZRS are placed at $\omega/\omega_{c}=j+1/4$ 
($j$ is integer) and ZRS are observed in a wide frequency range from 7.5 \cite{3} to 
240~GHz \cite{17}, while in bilayer systems ZRS develop from the strongest minima of combined 
MIS-MIRO oscillations and their location also depends on subband splitting.

\begin{figure}[ht]
\includegraphics[width=8cm]{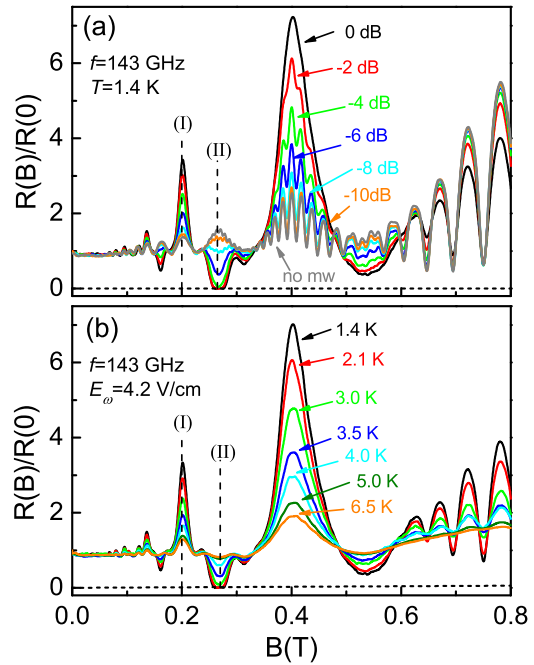}
\caption{\label{fig3} (color online) (a) Power dependence of magnetoresistance at a constant 
temperature (1.4~K) and (b) temperature dependence at a constant MW electric field (4.2~V/cm). 
With decreasing MW power, ZRS disappears at $-4$~dB and magnetoresistance for peak (II) 
approaches the dark value for $-10$~dB. With increasing $T$, ZRS also narrows and disappears.}
\end{figure}

In Fig. \ref{fig3} we present both power and temperature dependence of magnetoresistance at 
143~GHz, when ZRS develops around $B$=0.27~T with the largest width. In Fig. \ref{fig3}(a) we 
show power dependence at 1.4~K starting from 0~dB (highest intensity) to $-10$~dB, as well as the dark 
magnetoresistance (no MW). With decreasing MW power, the amplitudes of all enhanced and inverted 
MIS oscillations decrease while ZRS [peak (II)] first narrows and then disappears (at $-4$~dB) with 
decreasing MW power. The temperature dependence in Fig. \ref{fig3}(b) is carried out at a fixed 
MW power (0~dB, as in Fig. \ref{fig1}), which corresponds to a MW electric field $E_{\omega} \simeq
4.2$~V/cm. The effect of increasing temperature on the amplitudes of magnetoresistance oscillations 
is similar to the effect of decreasing power. The ZRS starts to narrow at $T=2.1$~K and disappears at 
$T \simeq 2.5$~K.

\begin{figure}[ht]
\includegraphics[width=9cm]{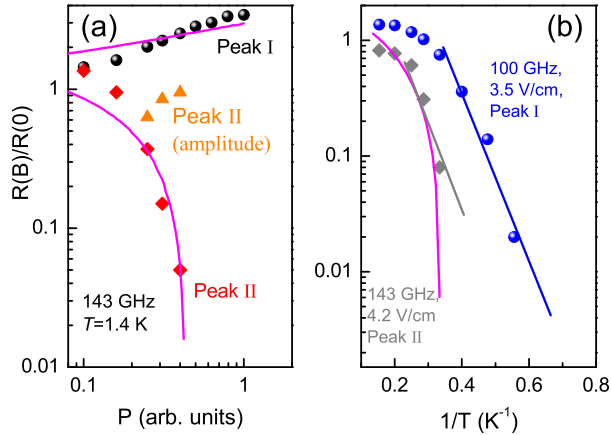}
\caption{\label{fig4} (color online) Analysis of power and temperature dependence. (a) The 
normalized magnetoresistance for peak (I) and peak (II), together with relative amplitude 
of peak (II) for 143~GHz at 1.4~K as a function of MW power. Peak (I) and the amplitude 
of peak (II) exhibit sublinear power dependence. The magnetoresistance for the inverted 
peak (II) strongly decreases with increasing MW power. (b) $R(B)/R(0)$ for 100~GHz 
[ZRS, peak (I)] and 143~GHz [ZRS, peak (II)]. The straight solid lines are fits to 
$R \propto\exp(-E_{ZRS}/T)$. The curves in (a) and (b) are theoretical plots.}
\end{figure}

Now we analyze power and temperature dependence of $R(B)/R(0)$ for both MIS peaks transformed into
ZRS. In Fig. \ref{fig4}(a) the magnitudes of the enhanced peak (I) and of the inverted peak 
which evolves into the ZRS (II) are plotted as functions of MW power. For the enhanced peak (I) 
$R_{xx}$ shows a sublinear power dependence before saturation. For the inverted MIS peak (II), 
$R_{xx}$ decreases very fast in a narrow power range. The amplitude of the inverted peak 
[i.e. $1-R(B)/R(0)$] scales similar to peak (I). Concerning temperature dependence, we observe 
a strong decrease of $R(B)/R(0)$ in the low-temperature region, when approaching ZRS. Therefore, 
we have constructed an Arrhenius plot which is presented in Fig. \ref{fig4}(b) for 143~GHz [ZRS for 
peak (II)] and 100~GHz [ZRS for peak (I)]. Applying the expression $R_{xx} \propto \exp(-E_{ZRS}/T)$, 
we obtain the activation energy $E_{ZRS}$=7~K, which is of the same order as in single-layer 
systems \cite{1,2,3}. For higher $T$, we see a deviation from the activation dependence, in a 
similar fashion as in Ref. 3. In the region where the resistivity is not very 
low [$R(B)/R(0)> 0.2$], the relative change of $R(B)/R(0)$ with temperature can also be fitted by 
the dependence $\propto -T^{-2}$, for both inverted peaks. 

The origin of ZRS is conventionally described in terms of the formation of a domain structure 
when the resistivity becomes absolutely negative. To check whether a negative 
resistivity can be reached in our sample under MW excitation, we have calculated the magnetoresistance 
based on both inelastic and displacement contributions. Calculations reveal that the displacement mechanism 
becomes essential only for higher MW electric fields and it is justified to neglect its contribution to 
photoresistance for our experimental conditions. In the regime of classically strong magnetic fields 
$\omega_c \tau_{tr} \gg 1$, where $\tau_{tr}$ is the transport time, and under a valid condition 
$\varepsilon_F \gg \Delta/2$, where $\varepsilon_F=\hbar^{2}\pi n_{s}/2m$ is the Fermi energy 
($\varepsilon_F \simeq 16.3$~meV for our sample), the magnetoresistance of the balanced double-layer 
system is expressed as
\begin{equation}
\frac{R(B)}{R(0)} = \int d \varepsilon \left(-\frac{\partial f_{\varepsilon}}{\partial \varepsilon} \right) 
{\cal D}^2_{\varepsilon}, 
\end{equation}
where $f_{\varepsilon}$ is the distribution function of electrons and ${\cal D}_{\varepsilon}$ is the 
dimensionless (expressed in units of $2 m/\pi \hbar^2$) density of states. The influence of microwaves 
on the distribution function is described by the kinetic equation \cite{8,18}
\begin{equation}
\frac{P_{\omega}}{4 \tau_{tr}} \sum_{\pm} {\cal D}_{\varepsilon \pm \hbar \omega} 
(f_{\varepsilon}-f_{\varepsilon \pm \hbar \omega}) = - \frac{f_{\varepsilon}-f^e_{\varepsilon}}{\tau_{in}},
\end{equation}
where the generation term (left-hand side) accounts for single-photon absorption of electromagnetic waves, 
and the electron-electron collision integral (right-hand side) is written through the inelastic relaxation 
time $\tau_{in}$. Next, $f^e_{\varepsilon}$ is the Fermi-Dirac distribution with effective electron 
temperature $T_e$ due to heating of the electron gas by microwaves and $P_{\omega} \propto E_{\omega}^2$ 
is the dimensionless MW power. The density of states for a two-subband system has been calculated 
numerically by using the self-consistent Born approximation. The quantum lifetime 
$\tau_{q} \simeq 7.1$~ps at 1.4~K ($\tau_{tr}/\tau_{q} \simeq 10$) is extracted from the 
amplitudes of MIS oscillations in dark magnetoresistance measurements. The temperature 
dependence of the inelastic relaxation time is taken in the usual form \cite{8} 
$\hbar/\tau_{in} = \lambda_{in} T_e^2/\varepsilon_F$, where $\lambda_{in}$ is a numerical constant 
of order unity. The electron temperature $T_e$, which depends on MW power, frequency, and magnetic field, 
is determined from the balance equation assuming energy relaxation of electrons due to their 
interaction with acoustic phonons. 

The result of calculations based on Eqs. (1)-(2) for 143~GHz and 0~dB attenuation (corresponding 
MW electric field $E_{\omega}=4.2$~V/cm) with $\lambda_{in}=0.5$ is shown in Fig. \ref{fig1}. 
The theoretical plot in Fig. \ref{fig1} reproduces all experimentally observed features 
and shows the negative resistivity around $B=0.27$~T [peak (II)]. The calculations for 100 GHz 
excitation ($E_{\omega}=3.5$~V/cm) also show the negative resistivity around $B=0.2$~T [peak (I)]. 
The power and temperature dependences of the resistance are in reasonable agreement with experiment; 
see Fig. \ref{fig4}. In particular, the disappearance of ZRS at $-4$~dB ($E_{\omega}=2.65$~V/cm) 
shown in Fig. \ref{fig3}(a) correlates with the results of theoretical calculations.  

The deviation of theory from experimental data is probably related to a limited applicability 
of the relaxation-time approximation \cite{8} in Eq. (2) and to a neglect of confined 
magnetoplasmon effects on MW absorption \cite{19}. Another source of deviation may be the 
assumption about homogeneous distribution of the MW field inside the sample and a neglect of the role of 
contacts or edges. Recently observed insensitivity of MIROs to the sense of circular 
polarization \cite{20}, the absence of MIROs in contactless measurements \cite{21}, and 
theoretical calculations of the MW field distribution in 2DES with metallic contacts \cite{22} 
indicate the near-contact origin of the phenomena of MIRO and ZRS in high-mobility layers. 
Since we have used linearly polarized microwaves and have not performed contactless 
measurements, we cannot definitely say whether a strong MW field gradient near the contacts 
is important for photoresistance in our bilayers. The absence of clear knowledge about the
specific nature of the near-contact oscillating contribution to resistivity \cite{22} 
does not allow us to discuss this problem in more detail, on the quantitative level. 

In conclusion, we have observed ZRS in a high-mobility bilayer electron system exposed to 
MW irradiation. The manifestation of ZRS in bilayers is distinct from the case of single-layer 
systems, due to a peculiar magnetotransport picture of a two-subband 2DES, where quantum 
magnetoresistance is modulated by MIS oscillations. Since the vanishing resistance 
develops for inverted MIS peaks, the ZRS intervals are narrow, and these states are very 
sensitive to MW frequency. We do not exclude the possibility that for samples with higher 
mobility two or more inverted MIS peaks could evolve into ZRS. The observed ZRS can be 
described as a consequence of current instability under absolute negative resistivity. 
By demonstrating experimentally that the ZRS phenomenon is not limited to single-layer 2DES, 
and, theoretically, that the presence of intersubband scattering in bilayers is not an obstacle 
for the negative resistivity, we suggest that similar physical mechanisms might lead to ZRS 
in multilayer or quasi-3D electron systems.
 	
This work was supported by CNPq, FAPESP, and with microwave facilities from ANR MICONANO.

\end{document}